\documentstyle[a4,12pt]{article} 

\input{epsf}		
\def\be{\begin{equation}}
\def\ee{\end{equation}}
\def\eref#1{(\ref{#1})}

\begin{document}
\bibliographystyle{unsrt}
\begin{titlepage}
{\hfill SWAT/75}
\vskip4cm
\centerline{\Large{\bf A Cluster Algorithm for the $Z_2$ Kalb-Ramond Model. }}
\vskip1.5cm
\centerline{\bf P.~K.~Coyle and I.~G.~Halliday}
\vskip0.8cm
\centerline{\it Department of Physics}
\centerline{\it University of Wales, Swansea}
\centerline{\it Singleton Park}
\centerline{\it Swansea, SA2 8PP, U.K. }
\vskip1.5cm

\abstract

	A cluster algorithm is presented for the $Z_2$
Kalb-Ramond plaquette model in four dimensions which
dramatically reduces critical slowing. The critical exponent
$z$ is reduced from $ z>2$ (standard Metropolis algorithm) to
$z= 0.32\pm0.06$. The Cluster algorithm updates the
monopole configuration known to be responsible for the 
second  order phase transition. 
\end{titlepage}

\section{Introduction}
	The standard local algorithms \cite{metrop} for updating
 monte-carlo simulations become very inefficient when applied to
 systems near a phase transition. As the critical point is approached 
 correlation lengths diverge signifying large scale structures on the
 lattice. These structures make simulations much more difficult.
 In order to  significantly change the relevant  objects, 
 lattice variables with large separations  have to be changed coherently. 

As the linear size (and therefore correlation lengths at criticality) 
is increased the algorithm has to be applied more times to achieve a
statistically independent configuration. Thus  as the lattice size
is increased not only will each update take longer but many more
updates will be needed.

Cluster algorithms try to address this problem by identifying the structures
relevant to the dynamics of the phase transition and using them as the
dynamical variables. In this way it is possible to construct an algorithm 
which acts directly on the relevant objects while still producing 
configurations with the correct statistical weight.

Cluster methods were first introduced for the Ising model 
\cite{swendsen,wolff1} and have since been extended to other spin models,
 in some cases reducing the critical exponent to near zero.
 Attempts to apply the same principles to gauge theories have been 
successful for some models such as the $Z_2$ gauge theory
in three dimensions \cite{sorin1}  and some vertex models \cite{evertz1}. 
However simulations of other gauge groups,
like  SU(3), near criticality remain plagued by Critical slowing down.

        The $Z_2$ plaquette theory \cite{Kalb_Ramond,Wegner}, being the centre of SU(2) is thought 
to be responsible for the phase transition seen in its adjoint representation, 
and therefore closely related to the second order point in the 
fundamental-adjoint phase diagram \cite{igh1,igh2}.

In \cite{sorin1} Ben-Av et.al. showed that the dynamics governing the phase
transition in a three dimensional $Z_2$ gauge theory  is governed by gauge
independent vortex loops and developed an algorithm to stochastically 
update these loops and thus dramatically reduce critical slowing down.
 Here we will show that these techniques can be extended to the 4
dimensional $Z_2$ plaquette theory where the dynamics are very similar.

\section{The Model}
	The dynamical  variables $P_{x,\mu\nu}$ are defined on the
 plaquettes of a 4 dimensional hyper-cubic lattice and can take on
 values $\pm 1$. The action is defined on 3 dimensional hypercubes
 $C$ and is given by:

\be
 {\rm S}={\sum_{C} {\rm S_c(\delta C) }} \label{eq:act}
\ee
where
\be
 {\rm S_c(\delta C)}={-\prod_{\delta C} P_{x,\mu\nu}}  
\ee
and  $\delta C$ are the plaquettes on the surface of the cube $C$. 

Each 3-cube has a value of $\pm 1$ and corresponds to a link on the 
dual lattice.
 Configurations can be uniquely defined, up to a gauge 
transformation by the links on the dual lattice. The bianchi identity for 
this model constrains the dual lattice configuration to be a system of
frustrated (-1) dual link loops embedded in a sea of satisfied (+1) dual links.
These dual loops are the world lines of monopoles \cite{Kalb_Ramond}.

This conserved monopole current $M_z^{\sigma}$ is given by
\be
M_z^{\sigma}=\epsilon^{\sigma\rho\mu\nu}\Delta_{\rho}P_{x,\mu\nu} \label{eq:mono}
\ee
with
\be
\Delta_{\sigma}M_z^{\sigma}=0 \label{eq:mono_wld}
\ee
 Here $M_z^{\sigma}$ lies on links of the dual lattice and z labels 
the dual site corresponding to the hypercube at x. 
Integrating $P_{x,\mu\nu}$ around a 3-cube gives
\be
M_z^{\sigma}=\oint P_{x,\mu\nu} \partial C 
\ee

Summing this over all 3-cubes (or dual links) gives the monopole world
line density which  is  equivalent to the Gibbs free energy.
 
	This model is dual to the 4 dimensional Ising model which has been
extensively studied \cite{Montavy_Weisz}  and gives the most accurate value
for the  $Z_2$ phase transition at $\beta=0.953$. 

\section{The Algorithm}
The algorithm used is a single cluster \cite{wolff1} update performed
 on the dual lattice
(ie on the 3-cubes) and is therefore gauge invariant. It follows the same
structure as the 3 dimensional case \cite{sorin1}.

\begin{figure}
\centerline{
\hbox{\epsfxsize=6cm \epsfbox{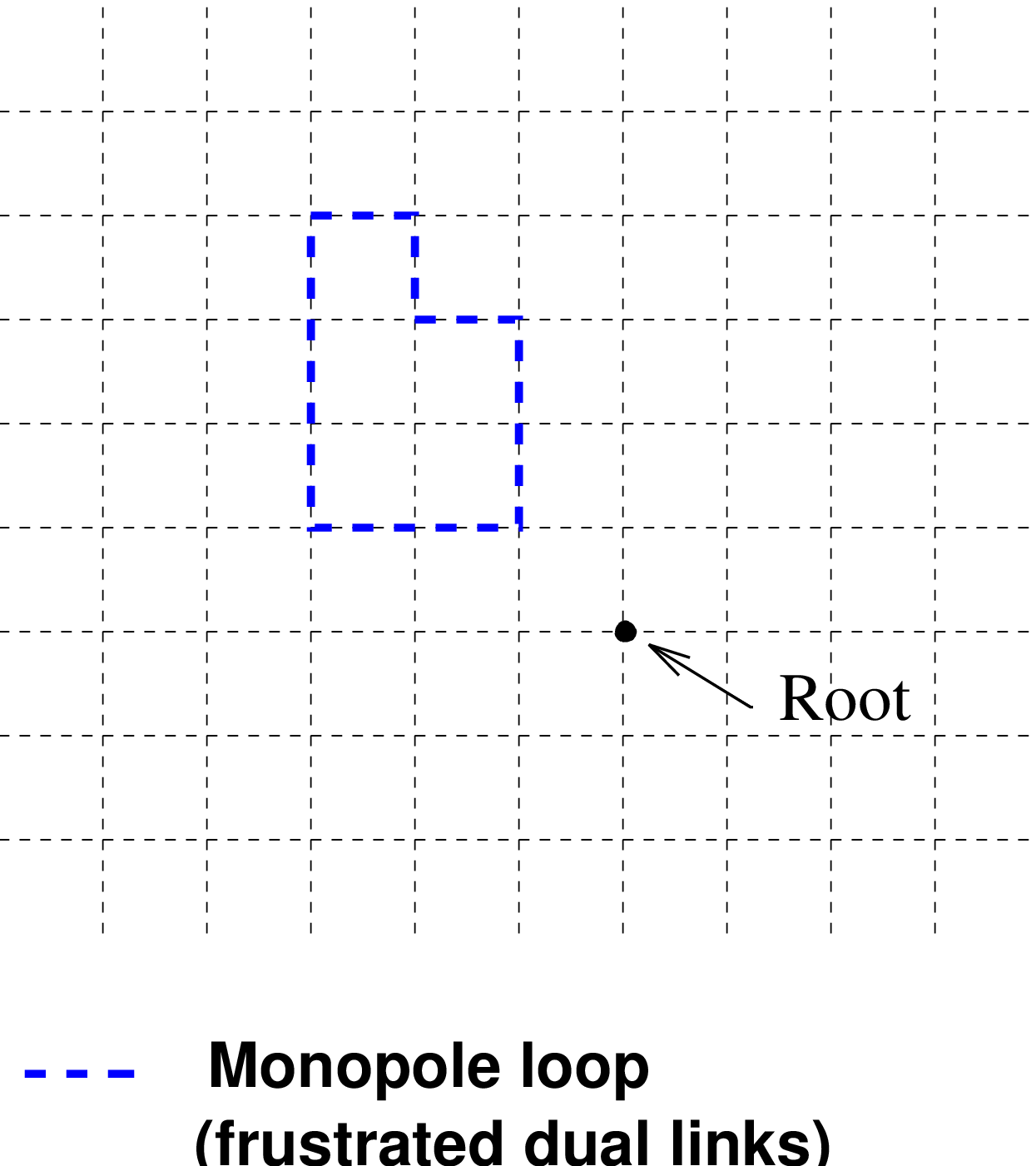}}
\hskip 1cm
\hbox{\epsfxsize=6cm \epsfbox{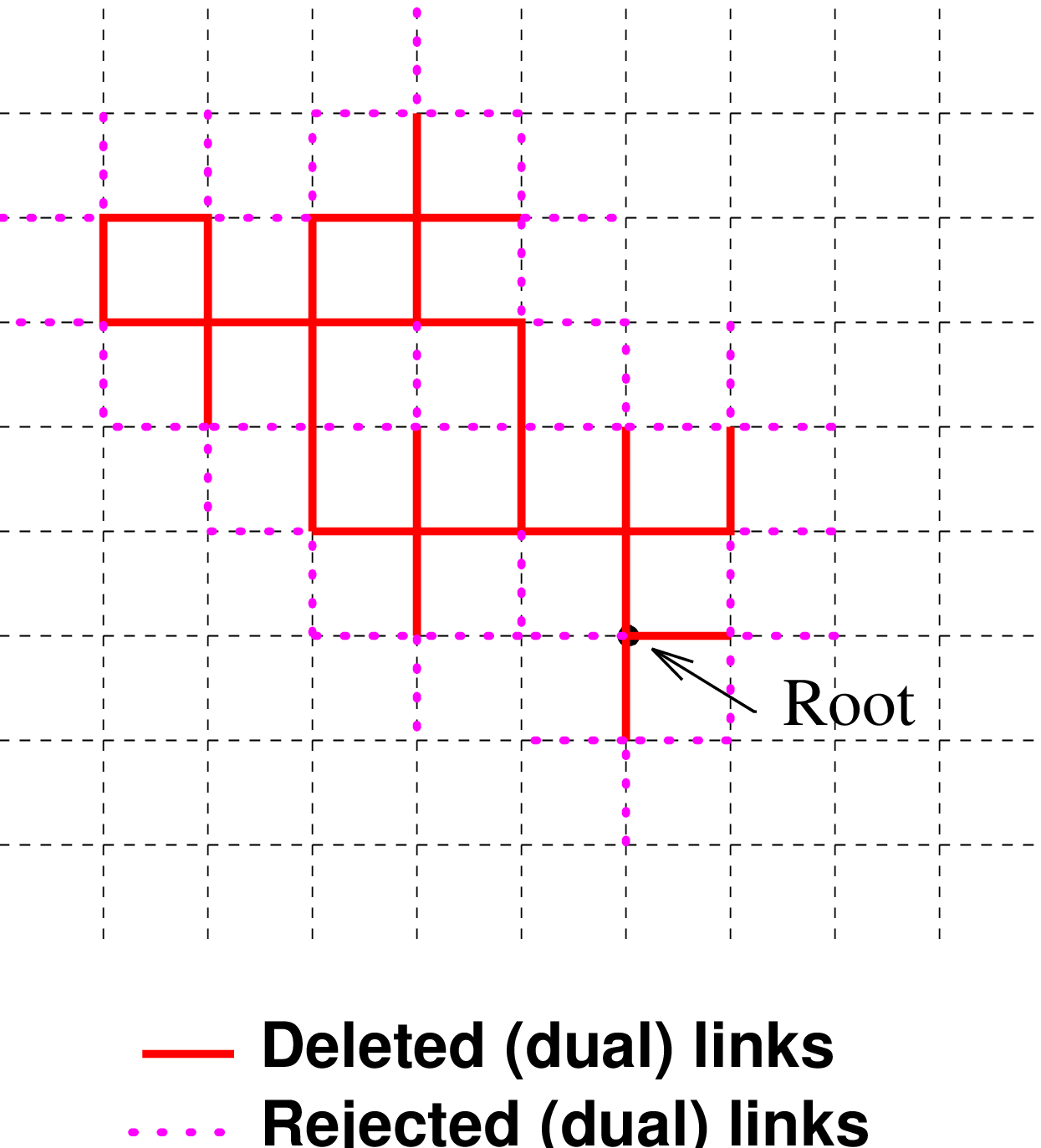}}
}
\vskip 1cm
\centerline{\epsfxsize=6cm \epsfbox{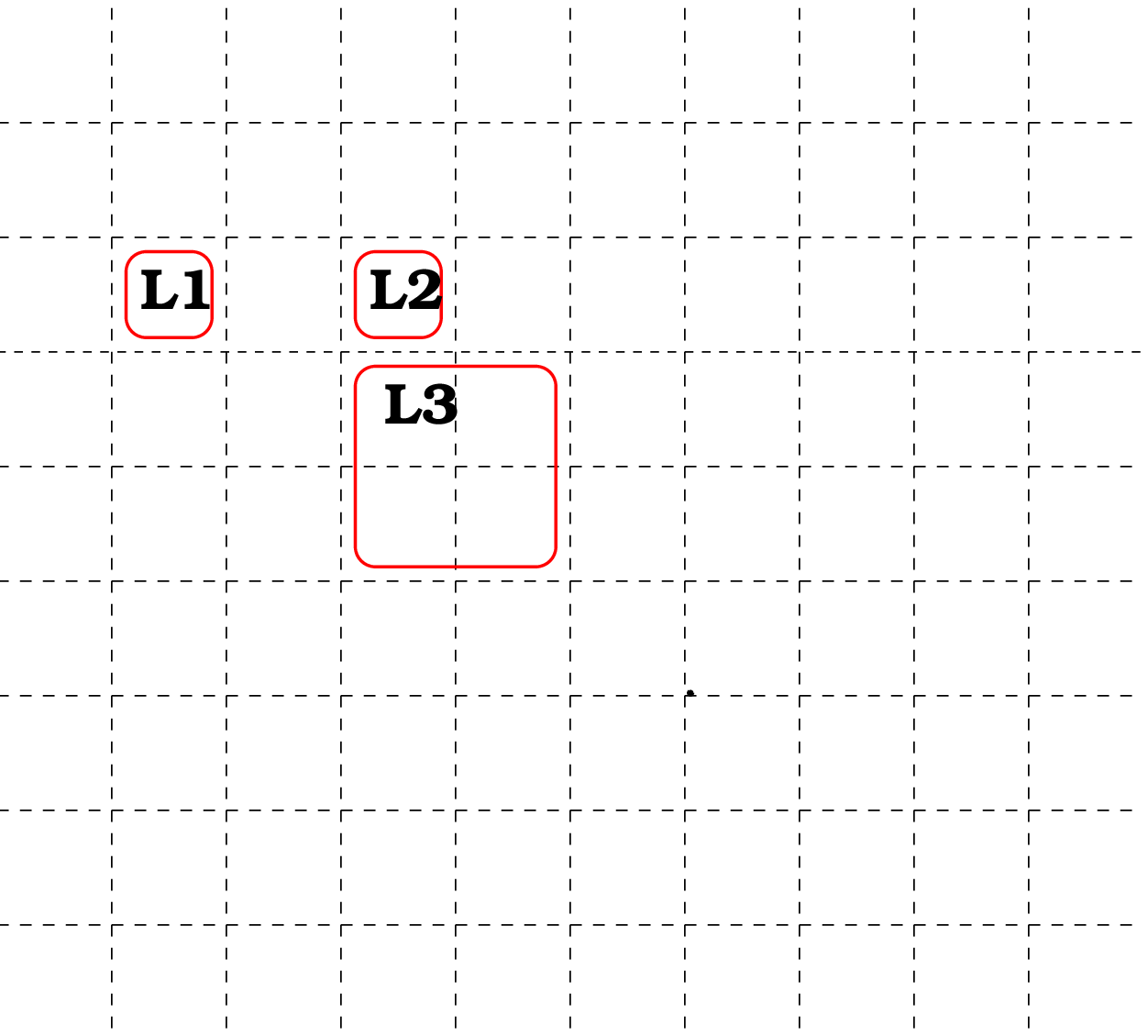}}	
\caption{a) The initial monopole configuration and a randomly picked
root. b) A graph of deletions built from the root. c) The three 
loops which can be updated independently. }
\label{fig:spanning_tree}

\end{figure}

The algorithm works by modifying the path taken by the monopole world
lines using the Boltzmann factor to weight any changes to the
configuration. A random starting point is chosen to build a tree of 
possible paths. This tree is called the graph of deletions.
 Fig.\ref{fig:spanning_tree} gives an example of how
such a tree would be constructed (the example has been restricted to 2
dimensions for clarity).
 Starting with the previous configuration pick a random point
to be the root (Fig.\ref{fig:spanning_tree}a). Visit each direction 
and add the neighbouring node  with probability

\be
P_d = e^{\beta(S_c-1)} \label{eq:prob}
\ee

Thus if an existing loop is encountered it is automatically added to
the graph of deletions. Non-frustrated dual links are added with a
probability dependent on the coupling $\beta$.
(Fig.\ref{fig:spanning_tree}b) shows a possible graph of deletions.
All dual links not in the graph of deletions are frozen.

\begin{figure}
\label{fig:spanning_tree2}
\centerline{\epsfxsize=10cm \epsfbox{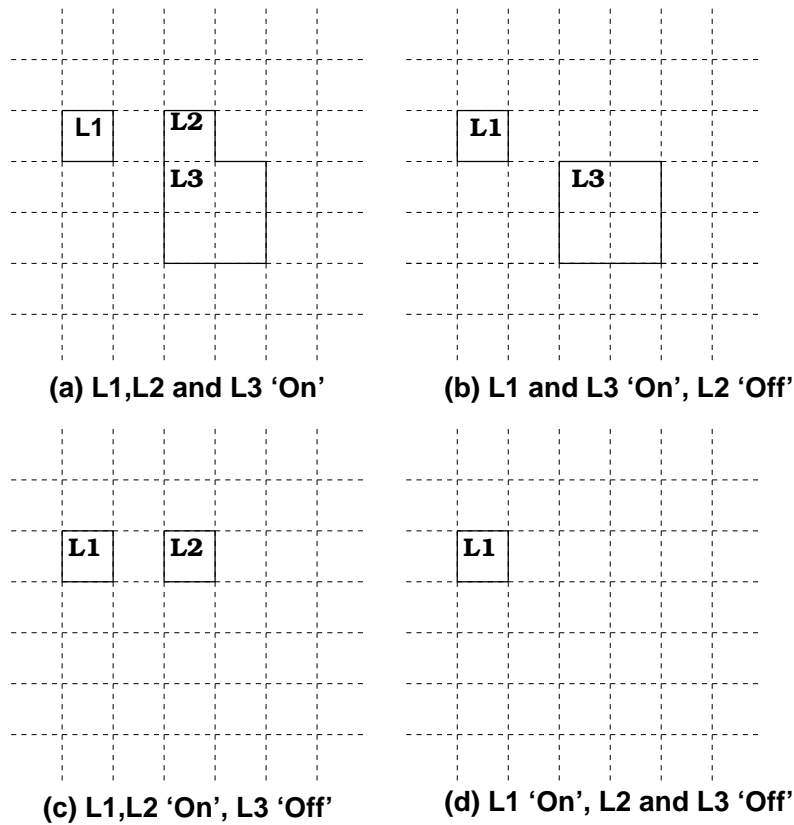}}	  
\caption{  4 of the 8  configurations possible from  this
graph of deletions. The other 4 have L1 turned off.  }
\end{figure}

The bianchi identity can be used to uniquely update the external legs
of the tree as valid configurations can only contain closed loops
 (Fig.\ref{fig:spanning_tree}c).
 The only degrees of
freedom left for changing the deleted links is a random choice 
of turning each loop on or off. That is choosing the links 
in each loop to be frustrated (on) or satisfied (off).
  Fig.\ref{fig:spanning_tree2}
shows 4 of the  new configurations which are possible by updating
the  graph of deletions. The remaining 4 possibilities have L1 
turned off. Each outcome is weighted equally.

Using the language of Kandel and Domany
\cite{kandel} the dual links forming the cluster are said to be
deleted and form a {\bf graph of deletions}. All other dual links are 
frozen. In this language the  graph of deletions is updated by
randomly  choosing a new monopole configuration which does not
change any of the frozen dual links.

It is easy to see how the algorithm can modify or even destroy an
existing monopole loop while also being able to create new loops of
arbitrary size.

The algorithm can easily be implemented on a computer as follows.
\begin{enumerate}
\item  Pick a random dual site (Hypercube) to be the root of the {\bf graph of 
       deletions}.
\item  Visit each dual link (3-cube) connected to this site and add it to the
       graph of deletions with probability $P_d = e^{\beta(S_c-1)}$.
       These links are said to be deleted. 
\item  Keep repeating the above step for each of the newly added dual
       sites until all of the dual links connected to the graph of deletions 
       have been visited  making sure that each dual link is only ever
       visited once.
\item  Identify all the loops within the graph of deletions and choose 
       one dual link for each loop to be a {\bf free link}. This leaves a 
       {\bf spanning tree}.
\item  Randomly assign a value to each of the free links. 
\item  All remaining links in the spanning tree are constrained so the
       bianchi identity can be used to update each link in the tree and 
       create a valid configuration.
\end{enumerate}

	The algorithm is obviously ergodic as there is a finite
probability of  deleting  all of the  dual links and choosing a
completely random configuration from the set of all valid configurations. 
Detailed balance also follows along the usual lines.

\section{Measurements and error analysis}

To measure how effective the algorithm is at producing 
independent configurations we calculated the normalised autocorrelation
function \eref{eq:nacf} for the Monopole density $M$.

\be
 A_M(t)= {{<M_sM_{s+t}>-<M>^2} \over{<M_s^2>-<M>^2} }  \label{eq:nacf}    
\ee

Integrating \eref{eq:nacf}  gives the integrated autocorrelation time
$\tau_{int}$ \eref{eq:iacf} .
\be
 \tau_{int,M}= {1\over2}\sum_{t=-\infty}^{+\infty} A_M(t)  \label{eq:iacf}
\ee
$\tau_{int}$ measures how many more updates are 
needed due to the configurations not being independent. 
For a sample mean:
\be
 \bar{M} = {1\over n}\sum_{t=1}^n M_t  
\ee  
the variance in $\bar{M}$ is

\be
Var(\bar{M}) = {1\over n}2\tau_{int}{(<M_s^2>-<M>^2)} 
\ee 
for $n>>\tau$. This is $2\tau_{int}$ times larger than expected for
independent configurations.

The Autocorrelation function  decays exponentially at least for 
large $s$. Fitting the tail of \eref{eq:nacf} to an exponential gives 
$\tau_{exp}$ the exponential autocorrelation time.
$\tau_{int}$ is defined so as to make $\tau_{int} \approx \tau_{exp}$
for large $\tau$.
The autocorrelation time  diverges as a power of the correlation
length so at the critical point $\tau$  varies with the lattice
 size L as:

\be
 \tau \approx L^z \label{eq:critexp} 
\ee

The critical exponent $z$ quantifies the extent to which critical
slowing down is affecting the simulation. Local algorithms tend to
have $z\approx2$ for models with a phase transition.

\section{Results}

	Simulations were performed on Dec 3400AXP Workstations.
For comparison a standard Metropolis algorithm was also studied.
 The monopole density \eref{eq:mono}  was calculated after each update.
  A number of simulations
 were performed for $\beta$ values around the critical point in order to
find the largest autocorrelation time
 for the lattice size. Both exponential ($\tau_{exp}$)
and integrated ($\tau_{int}$) autocorrelation time were measured for the
Monopole  density. Both were found to be in agreement although 
for the cluster algorithm  $\tau_{exp}$ proved easier to calculate
 as the autocorrelation functions fitted a exponential very closely
 before the tail became too noisy.
  Errors in the autocorrelation function were calculated
 using blocking.
 We noticed auto-correlation functions for metropolis were not pure
 exponentials. To fit this to an exponential decay it was  necessary
 to find a window in the autocorrelation function after an initial
 power law dependence but before noise takes over. Metropolis  gave
 a critical exponent of $z_c>2$ as expected for a local algorithm. 

As our algorithm only updates one cluster at a time the effect of each 
application can vary  from updating no links to updating the entire lattice. 
In order to make a useful comparison of the CPU requirements for each
algorithm the sweeps of the cluster algorithm were scaled by a factor 
$C$ representing the  average cluster size.
This factor was calculated as follows.
Each application of the algorithm is said to be one hit. 
This hit touches $C$ dual links of the lattice.
$S$ of these links form the spanning tree. 
$F$ are free links.
$R$ links are looked at but rejected.
The graph of deletions therefore contains $S+F$ links.
Fig.1 shows the CPU load  is roughly proportional to the cluster size
$C$ which we have defined  as   $C=S+F+R$.

\begin{figure}
\centerline{\epsfxsize=12cm \epsfbox{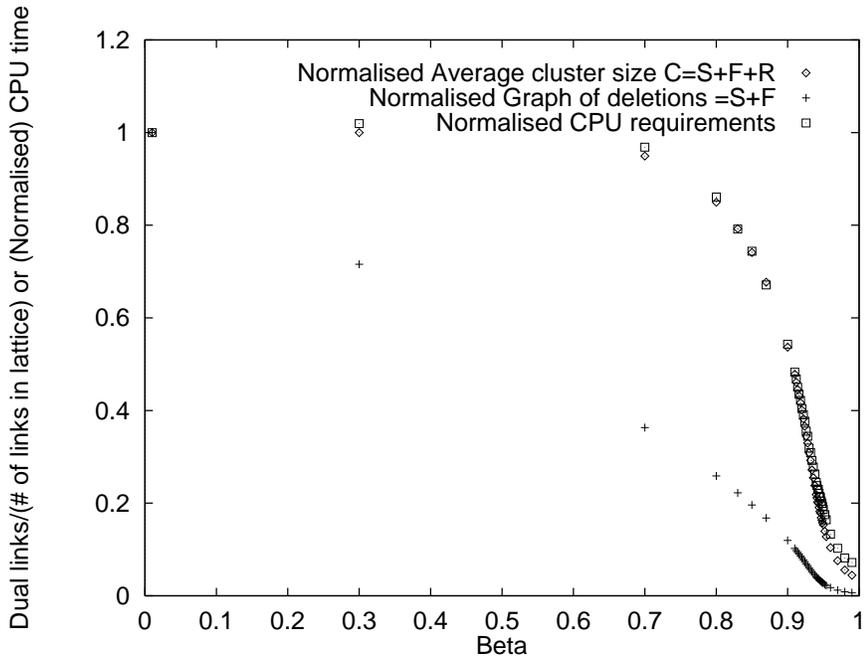}}	  
\caption{A comparison of Cluster sizes and CPU requirements for a
$5^4$ lattice}
\end{figure}

One sweep of the lattice is defined as $N/C$ hits where $N$ is the total
number of links on the lattice. Thus a sweep takes 
approximately the same time for all $\beta$ and each dual link is touched
on average once per sweep.
Each Cluster sweep took approximately the same time as a metropolis
sweep. The extra complexity is compensated for by working on the
smaller dual lattice where it is easier to calculate the action.
Our code for metropolis updated approximately $0.42\times10^6$
plaquettes/s a second while the cluster algorithm updates approximately
$0.3\times10^6$ dual links/s.

Runs consisted of $1\times10^6$ --- $2\times10^6$ clusters which gives
 approximately $6\times10^4$ --- $6\times10^5$ sweeps. 
Fig \ref{fig:act} shows the results for the autocorrelation time
 $\tau_{int}$ against lattice size.
The solid lines are our best fits of $z_c=0.32\pm0.06$ for the cluster
algorithm and $z_c=2.58\pm0.1$ for metropolis.

\label{fig:act}
\begin{figure}
\centerline{\epsfxsize=12cm \epsfbox{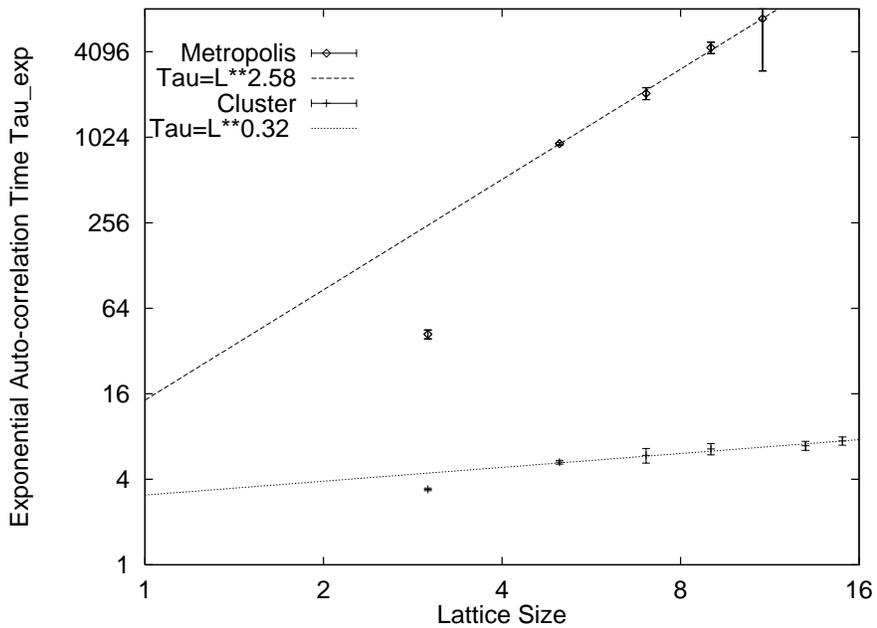}}	  
\caption{ Monte-carlo simulations of $Z_2$ Kalb-Ramond
model in 4D using Metropolis and Cluster algorithms. Autocorrelation
time vs Lattice size. }
\end{figure}

\section{Conclusions}
	The Cluster algorithm presented above becomes more efficient 
 than Metropolis for lattice sizes as small as $4^4$ and is
approximately 1000 times more efficient for a $15^4$ lattice.
Comparing our algorithm to Wolff's for the 4D Ising model 
gives further  support to  the idea of universality classes.
You can see from Table~\ref{table::table1} that dynamical
 critical exponents
for the $Z_2$ models and their corresponding dual Ising  models
 are the same within errorbars.

The embedding of Ising spins has been used  successfully to extended
the  realm of 
cluster  algorithms to other spin systems although not all embeddings
 work \cite{embed1}.   
It is hoped that embedding  this $Z_2$ algorithm in a
full $SU(2)$ theory 
will give some insight into the structure of the 
$SU(2)$ fundamental-adjoint phase diagram
\cite{igh1,igh2,IGH+LC+AS,finite_temp1,finite_temp2}
and the bulk transition
of pure SU(2).

\begin{table}
\begin{center}
\begin{tabular}{||c|c|c||}                     \hline
Model&  Swendsen--Wang &  Single Cluster \\ \hline \hline
$Z_2$ 4D    & --- & $0.32 \pm 0.06$  \\ \hline 
Ising 4D    & $0.83 \pm0.01$  & $0.29\pm0.07$  \\ \hline \hline
$Z_2$ 3D    & $0.73 \pm 0.06$  &  --- \\ \hline
Ising 3D    & $0.75\pm0.01$  & $0.39 \pm 0.01$   \\ \hline

\end{tabular}
\caption{Dynamical Critical Exponents for $Z_2$ models and the 
Ising models to which they are dual. }
\label{table::table1}
\end{center}
\end{table}

\end{document}